\documentclass[11pt,a4paper]{article}
\pdfoutput=1
\usepackage{soul}
\usepackage{jheppub}
\usepackage{amsfonts}
\usepackage{amsmath}
\allowdisplaybreaks[4]     
\usepackage{amssymb}   
\usepackage{euscript}       
\usepackage{color}
\usepackage[normalem]{ulem}
\usepackage{mathrsfs}  
\usepackage{booktabs}
\usepackage{siunitx}
\usepackage{amsmath,latexsym,xcolor,mathtools}
\usepackage{graphicx,verbatim, subcaption, caption,inputenc}
\usepackage{hyperref}

\begin{document}

\title{Weyl--invariant scalar--tensor gravities from purely metric theories}

\author[1,2]{Giorgos Anastasiou,}
\author[1,2,3]{Ignacio J. Araya}
\author[4]{and Avik Chakraborty}

\emailAdd{ganastasiou@unap.cl}
\emailAdd{ignaraya@unap.cl}
\emailAdd{avikchak@usc.edu}

\affiliation[1]{Instituto de Ciencias Exactas y Naturales, Universidad Arturo Prat, Playa Brava 3256, 1111346, Iquique, Chile}
\affiliation[2]{Facultad de Ciencias, Universidad Arturo Prat, Avenida Arturo Prat Chac\'on 2120, 1110939, Iquique, Chile}
\affiliation[3]{School of Mathematics and Hamilton Mathematics Institute, Trinity College, Dublin 2, Ireland}
\affiliation[4]{Departamento de Ciencias F\'isicas, Facultad de Ciencias Exactas, Universidad Andres Bello,
Sazi\'e 2212, Piso 7, Santiago, Chile}

\abstract{
We describe a method to generate scalar--tensor theories with Weyl symmetry, starting from arbitrary purely metric higher derivative gravity theories. The method consists in the definition of a conformally--invariant metric $\hat{g}_{\mu \nu}$, that is a rank (0,2)--tensor constructed out of the metric tensor and the scalar field. This new object has zero conformal weight and is given by $\phi^{2/\Delta}g_{\mu \nu}$, where $(-\Delta)$ is the conformal dimension of the scalar. As $g_{\mu \nu}$ has conformal dimension of 2, the resulting tensor is trivially a conformal invariant. Then, the generated scalar--tensor theory, which we call the Weyl uplift of the original purely metric theory, is obtained by replacing the metric by $\hat{g}_{\mu \nu}$ in the action that defines the original theory. This prescription allowed us to  define the Weyl uplift of theories with terms of higher order in the Riemannian curvature. Furthermore, the prescription for scalar--tensor theories coming from terms that have explicit covariant derivatives in the Lagrangian is discussed. The same mechanism can also be used for the derivation of the equations of motion of the scalar--tensor theory from the original field equations in the Einstein frame. Applying this method of Weyl uplift allowed us to reproduce the known result for the conformal scalar coupling to Lovelock gravity and to derive that of Einsteinian cubic gravity. Finally, we show that the renormalization of the theory given by the conformal scalar coupling to Einstein--Anti--de Sitter gravity originates from the Weyl uplift of the original renormalized theory, which is relevant in the framework of \textit{conformal} \textit{renormalization}.}
\maketitle

\section{Introduction}
\label{sec:intro}



In recent times, conformally invariant gravity theories have attracted a lot of attention in different branches of physics and mathematics. These theories are invariant under Weyl transformations, i.e., local scale transformations of the metric, $g_{\mu \nu} \rightarrow e^{2 \sigma(x)} g_{\mu \nu}$, that preserve the angles and not the distances. Their applicability spans a wide range of topics, e.g. in renormalization group (RG) flows, string theory, critical phenomena etc. Recently, it has been pointed out the significance that bulk Weyl symmetry has in the construction of gravity theories with well defined asymptotics, namely, free of IR divergences, that being one of the motivations of this paper. The simplest example of local scale invariance in the context of gravity is given by the four--dimensional Conformal Gravity (CG)~\cite{Weyl:1918ib,Weyl:1919fi,Bach:1921}, that admits Einstein gravity as part of its solution space\footnote{For a selection of references on CG, see Refs.~\cite{Bonora:1985cq,Deser:1993yx,Erdmenger:1997gy,Boulanger:2004zf,Bastianelli:2000rs,Metsaev:2010kp,Oliva:2010zd,Lu:2013hx,Lu:2011zk,Lu:2012xu,Maldacena:2011mk,Anastasiou:2016jix}. See also earlier works in Refs.~\cite{Tomboulis:1984dd,Julve:1978xn,Fradkin:1981iu}.}. Though CG is a higher derivative theory that allows ghosts (modes with negative kinetic energy), it has better renormalizability and UV properties than the Einstein gravity~\cite{Stelle:1976gc,Capper:1975ig,Julve:1978xn,Fradkin:1981iu} and hence it is considered a useful toy model for quantum gravity.

Now, modified theories of gravity have been subjects of interest over a long time, the main reason being that General Relativity (GR) is not renormalizable and hence can not be traditionally quantized. Interestingly, it turned out that the renormalization at one--loop requires the Einstein--Hilbert (EH) action to be corrected by terms of higher order in the Riemannian curvature~\cite{Utiyama:1962sn,Stelle:1976gc}. Also, it was realized that the effective low energy gravitational action demands higher order curvature invariants as well, when quantum corrections are taken into account~\cite{Birrell:1982ix,Buchbinder:1992rb,Vilkovisky:1992pb}. Along this line of constructing modified gravity theories, one of the most well--known alternatives to GR, is given by scalar--tensor theory~\cite{Bergmann:1968ve,Brans:1961sx,Dicke:1961gz,Faraoni:2004pi,Nordtvedt:1970uv,Wagoner:1970vr}.

In this context, one interesting extension, which is an active field of research, is that of conformal couplings of scalar fields to gravity theories, or in other words, the construction of Weyl--invariant scalar--tensor theories, which is the main focus of this paper. A notable feature of this class of theories is the presence of black holes with either primary or secondary hair, that motivated a great amount of work on the subject~\cite{Bocharova:1970skc,Bekenstein:1974sf,Bekenstein:1975ts,Xanthopoulos:1992fm,Cisterna:2021xxq}. Important work was done by Oliva and Ray in Ref.~\cite{Oliva:2011np} in the context of generalized Lovelock gravity, where they constructed a four--rank tensor involving the curvature and the field derivatives, that transforms covariantly under Weyl rescalings, and which was used to construct the scalar conformal couplings to Lovelock theory.

Furthermore, the significance of bulk Weyl symmetry in the defintion of geometric objects with well--defined asymptotics in metric theories has been signaled in a series of papers~\cite{Grumiller:2013mxa,Anastasiou:2016jix,Anastasiou:2020mik}. In the same context, the authors in Ref.~\cite{Anastasiou:2022wjq}, have shown that  the requirement of full local Weyl invariance for the bulk fields for a conformally coupled scalar--tensor theory with quartic potential, renders the action finite. In this paper, we will present how to construct a Weyl invariant scalar--tensor theory of gravity starting from any diffeomorphism--invariant purely metric action. Our proposal will be supported by a series of examples, including the cases of Einstein, Lovelock and other  higher curvature gravity (HCG) theories. 

The paper is organized as follows. In section~\ref{sec:recipe}, we present the prescription that will allow us to promote any purely metric Lagrangian to a Weyl invariant scalar--tensor theory. In section~\ref{sec:wu-eads}, we will provide a trivial example of our method for the EH--Anti--de Sitter (AdS) case in generic $D$ dimensions. In section~\ref{sec:wuhct}, we will show how this prescription can be 
applied in order to derive the most generic scalar conformal couplings to higher curvature gravity theories and provide examples that include Lovelock theories (section~\ref{subsec:lovelock}) 
and quasitopological gravity theories (section~\ref{subsec:qtg}). In section~\ref{sec:wu-ren}, we will discuss about the relation between Weyl symmetry and renormalization through the concrete example of Einstein--AdS action written in MacDowell--Mansouri form in $D=4$, being Weyl--uplifted following our prescription. Section~\ref{sec:discuss} will be devoted for the summary of our results and future developments.

\section{Scalar--Tensor Weyl Uplift of Metric Theories}
\label{sec:recipe}


In this section, we present a general prescription for constructing Weyl--symmetric scalar--tensor theories of gravity, starting from an arbitrary purely metric diffeomorphism--invariant theory, which to the best of our knowledge has not been discussed before.

The prescription considers purely--metric gravity actions of arbitrarily high order in derivative and Riemannian curvature, of the form

\begin{equation}
I = \int d^D x \sqrt{-g} \mathcal{L}\Big(R_{\rho \sigma}^{\mu \nu},\nabla^{\lambda} R_{\rho \sigma}^{\mu \nu},\ldots,\nabla^{\lambda_1} \cdots \nabla^{\lambda_n} R_{\rho \sigma}^{\mu \nu},g_{\mu \nu}\Big) \,. \label{eqn:gen-action-pure}
\end{equation}
Considering the local Weyl transformations of the metric and the scalar field

\begin{equation}
g_{\mu \nu} \rightarrow e^{2 \sigma} g_{\mu \nu} \ , \ \phi \rightarrow e^{-\Delta \sigma}\phi \ ,
\label{scalings}
\end{equation}
applied in the arbitrary diffeomorphism--invariant action above, we construct Weyl--invariant scalar--tensor gravity theories of the same order in derivative for a fixed scaling dimension of the scalar field.

In particular, the conformal dimension of the scalar field is fixed by requiring the global scale invariance of the canonical kinetic term, leading to the value

\begin{equation}
\Delta =\frac{D-2}{2} \,.\label{eqn:conf-dim-scalar}
\end{equation}
Then, for the scalings given above \eqref{scalings}, one defines the conformal invariant metric as

\begin{equation}
\hat{g}_{\mu \nu}=\phi^{2/\Delta}g_{\mu \nu} \,. \label{eqn:conf-inv-metric}
\end{equation}
Indeed, as the metric tensor has conformal weight $2$ in every dimension, the resulting conformal invariant metric $\hat{g}_{\mu \nu}$ is a rank--$(0,2)$ tensor which is unchanged under Weyl symmetry. Therefore, any tensor constructed out of the $\hat{g}_{\mu \nu}$ will be Weyl invariant, that will be the guiding principle for our analysis.\footnote{Note that the method of introducing a gauge symmetry by adding a compensating field is known as the Stuckelberg trick, which was used in Ref.~\cite{Komargodski:2011vj} to construct the effective dilaton action in 4D.}

In particular, the prescription that will be used for the derivation of a Weyl--invariant scalar--tensor theory, is based on the replacement of the original metric $g_{\mu \nu}$ by $\hat{g}_{\mu \nu}$ in the action \eqref{eqn:gen-action-pure}. Explicitly, this class of theories has the generic form given by

\begin{equation}
I[\hat{g}_{\mu \nu}] = \int d^D x \sqrt{-\hat{g}} \mathcal{L}\Big( \hat{R}_{\rho \sigma}^{\mu \nu},\hat{\nabla}^{\lambda} \hat{R}_{\rho \sigma}^{\mu \nu},\ldots,\hat{\nabla}^{\lambda_1} \cdots \hat{\nabla}^{\lambda_n} \hat{R}_{\rho \sigma}^{\mu \nu},\hat{g}_{\mu \nu} \Big) \,, \label{eqn:wu-gen-action}
\end{equation}
where $\hat{R}_{\rho \sigma}^{\mu \nu}$ and $\hat{\nabla}_{\lambda}$ are the Riemann tensor and the covariant derivative, respectively, computed with the conformal invariant metric. One important conceptual clarification is that, in the action of Eq.~\eqref{eqn:wu-gen-action}, the hatted quantities are $\emph{not}$ the metric, the Riemann tensor and the covariant derivative, but rather they are tensors and differential operators with the same functional forms as the un--hatted objects, but where instead of using the metric tensor to define them, one uses $\hat{g}_{\mu \nu}$. Namely, the Weyl uplift consists on maintaining the functional form of the Lagrangian density while replacing all the geometric objects by their hatted counterparts.  

In order to see more explicitly how the prescription works for Weyl-uplifting an arbitrary metric theory, which may contain explicit covariant derivative terms, contractions of the Riemannian curvature to arbitrary order, explicit dependence on the metric tensor and combinations thereof, it is convenient to examine the uplift of the Riemann tensor and the covariant derivative separately.

Firstly, we consider the Riemann tensor evaluated on the conformally invariant metric. From the transformation laws given in Ref.~\cite{Wald:1984rg}, we derive the identity that

\begin{equation}
\hat{R}_{\lambda \rho}^{\mu \nu}=R^{\mu \nu}_{\lambda \rho}\Big[\phi^{2/\Delta}g_{\mu\nu} \Big]=\phi^{-D/\Delta}S^{\mu \nu}_{\lambda \rho} \,, \label{eqn:rhat-sourya-tensor}
\end{equation} 
where $S^{\mu \nu}_{\lambda \rho}$ is defined in Ref.~\cite{Oliva:2011np} as

\begin{equation}
S^{\mu \nu}_{\lambda \rho}=\phi^2 R^{\mu \nu}_{\lambda \rho} - \frac{4}{\Delta}\phi \delta^{[\mu}_{[\lambda}\nabla_{\rho ]}\nabla^{\nu ]}\phi + \frac{4(\Delta +1)}{\Delta^2}\delta^{[\mu}_{[\lambda}\nabla_{\rho ]}\phi \nabla^{\nu ]}\phi - \frac{1}{\Delta^2}\delta^{\mu \nu}_{\lambda \rho}\nabla_\kappa \phi \nabla^\kappa \phi \,, \label{eqn:sourya}
\end{equation}
and its partial and full traces are given by

\begin{equation}
S^{\mu}_{\nu}=\delta^{\lambda}_{\rho }S_{\nu \lambda}^{\mu \rho}=\phi^2 R^{\mu}_{\nu} - \frac{1}{\Delta}\delta^{\mu}_{\nu}\nabla_\kappa \phi \nabla^\kappa \phi -\frac{\phi}{\Delta}\Box \phi \delta^{\mu}_{\nu}-2\phi \nabla_{\nu} \nabla^{\mu}\phi+\frac{2(\Delta+1)}{\Delta}\nabla_{\nu}\phi \nabla^{\mu}\phi \,, \label{eqn:sourya-1st-trace}
\end{equation}
and

\begin{equation}
S=\frac{1}{2}\delta^{\mu \nu}_{\rho \lambda}S_{\mu \nu}^{\lambda \rho}=\phi^{2}R-\frac{4(D-1)}{(D-2)} \phi \Box \phi \,,
\label{eqn:sourya-2nd-trace}
\end{equation}
respectively. On the same footing, the Levi--Civita connection of the conformally invariant metric reads

\begin{equation}
\hat{\Gamma}^\mu_{\nu \lambda} = \Gamma^\mu_{\nu \lambda} + \Theta^\mu_{\nu \lambda} \,, \label{eqn:wu-connection}
\end{equation}
where

\begin{equation}
\Theta^\mu_{\nu \lambda} = \frac{1}{\Delta \phi}\Big(\delta^\mu_\lambda \partial_\nu \phi + \delta^\mu_\nu \partial_\lambda \phi - g_{\nu \lambda}g^{\mu \sigma}\partial_\sigma \phi \Big) \,. \label{eqn:wu-connection-correction}
\end{equation}
Then, when computing the Weyl--uplifted covariant derivative of a tensor, one has to use the standard rules for the covariant derivative, but considering the modified connection, i.e.,

\begin{equation}
\hat{\nabla}_\nu = \nabla_\nu (\hat{\Gamma}^\mu_{\nu \lambda}) \,. \label{eqn:wu-covD}
\end{equation}
For example, its action on contravariant and covariant vectors, obtains the form

\begin{equation}
\hat{\nabla}_\nu v^{\mu} = \partial_\nu v^{\mu}+\hat{\Gamma}^\mu_{\nu \lambda}v^{\lambda}=\nabla_\nu v^{\mu}+\Theta^\mu_{\nu \lambda}v^{\lambda} \,, \label{eqn:wu-contra}
\end{equation}
and

\begin{equation}
\hat{\nabla}_\nu w_{\mu} = \partial_\nu w_{\mu}-\hat{\Gamma}^\lambda_{\nu \mu}w_{\lambda}=\nabla_\nu w_{\mu}-\Theta^\lambda_{\nu \mu}w_{\lambda} \,, \label{eqn:wu-covar}
\end{equation}
respectively.

Analogously, the equations of motion of the Weyl--uplifted theory can be obtained directly from those of the original theory, replacing the metric by the conformal--invariant metric and evaluating the Riemannian curvatures and covariant derivatives using $\hat{g}_{\mu \nu}$ with the explicit dependence on the scalar. The derivation is straightforward due to the relation between  $g_{\mu \nu}$ and $\hat{g}_{\mu \nu}$, as shown in Eq.~\eqref{eqn:conf-inv-metric}. In particular, starting from the action

\begin{equation}
I\left[\hat{g}_{\mu \nu}\right] = I\left[g_{\mu \nu},\phi \right] =\int d^D x \sqrt{-\hat{g}} \mathcal{L}\Big(\hat{R}_{\rho \sigma}^{\mu \nu},\hat{\nabla}^\lambda \hat{R}_{\rho \sigma}^{\mu \nu},\hat{g}_{\mu \nu}\Big) \,,
\end{equation}
we obtain the relation between the equations of motion (EOM) of the metric field in the Weyl and in the Einstein frame as

\begin{equation}
\frac{\delta I}{\delta g_{\mu \nu}} = \frac{\delta I}{\delta \hat{g}_{\sigma \tau}}\frac{\delta \hat{g}_{\sigma \tau}}{\delta g_{\mu \nu}} = \phi^{2/\Delta} \frac{\delta I}{\delta \hat{g}_{\mu \nu}} \,,
\label{eomg}
\end{equation}
or equivalently

\begin{equation}
\epsilon^{\mu \nu} \left(g_{\lambda \rho },\phi \right)=\phi^{2/\Delta} \hat{\epsilon}^{\mu \nu}\left(\hat{g}_{\lambda \rho}\right) \,,
\label{epsilong}
\end{equation}
where $\epsilon^{\mu \nu} \left(g_{\lambda \rho },\phi \right)=\frac{\delta I}{\delta g_{\mu \nu}}$. Here, $\hat{\epsilon}_{\mu \nu}\left(\hat{g}_{\lambda \rho}\right)$ describes field equations of the conformally invariant metric tensor in the Einstein frame, but in the Weyl frame it encodes a certain combination of the scalar field $\phi$ and the physical metric $g_{\mu \nu}$, as $\hat{\epsilon}^{\mu \nu}\left(\hat{g}_{\lambda \rho}\right)=\hat{\epsilon}^{\mu \nu}\left(\phi^{2/\Delta}g_{\lambda\rho}\right)$. In a similar way, for the EOM of the scalar field we get

\begin{equation}
\frac{\delta I}{\delta \phi} = \frac{\delta I}{\delta \hat{g}_{\mu \nu }}\frac{\delta \hat{g}_{\mu \nu }}{\delta \phi} = \frac{2}{\Delta \phi} \hat{\epsilon} \left(\hat{g}_{\lambda\rho}\right) \,,
\label{eomscalar}
\end{equation}
where $\hat{\epsilon} \left(\hat{g}_{\lambda\rho}\right)=\hat{g}_{\mu \nu} \hat{\epsilon}^{\mu \nu}\left(\hat{g}_{\lambda\rho}\right)$ is the trace of the EOM tensor in the Einstein frame. Interestingly enough, taking into account Eq.~\eqref{epsilong} and the definition $\hat{g}_{\mu \nu}=\phi^{2/\Delta} g_{\mu \nu}$, we get that

\begin{equation}
\hat{\epsilon} \left(\hat{g}_{\lambda\rho}\right)=\hat{g}_{\mu \nu} \hat{\epsilon}^{\mu \nu} \left(\hat{g}_{\lambda\rho}\right) = g_{\mu \nu} \epsilon^{\mu \nu} \left(\phi^{2/\Delta}g_{\lambda\rho} \right)=\epsilon \left(g_{\lambda \rho},\phi \right) \,.
\label{scalartrace}
\end{equation}
Defining the energy--momentum tensor of the scalar sector as the variation of the non--minimally--coupled scalar action with respect to the metric ($T_{\mu \nu}=\epsilon_{\mu \nu} \left(g_{\lambda \rho},\phi \right)$), the EOM of the metric tells us that the scalar is a stealth, i.e., $T_{\mu \nu}=0$. Moreover, the EOM of the scalar is given by the trace of $T_{\mu \nu}$, which vanishes identically due to the stealth condition. This results is compatible with the tracelessness requirement that any conformally invariant theory should satisfy. 

The previous expressions constitute the building blocks for obtaining the conformal scalar couplings to any metric theory, and in the following sections, we will study different interesting examples.

It may be argued that the theories of the form of Eq.~\eqref{eqn:wu-gen-action} are trivial, or that they have fake Weyl symmetry, as 
the scalar degree of freedom is simply the result of a field redefinition, and it is a redundant degree of freedom that can be gauged away by fixing the Weyl frame to be the frame where the metric is equal to the conformally invariant metric. Nevertheless, in cases when the boundary condition for the spacetime is fixed ---e.g., having a Fefferman--Graham expansion near the AdS boundary--- the physical metric is the one associated with the boundary behavior, which is the case in holography. Furthermore, for probes that couple to the geometry in the scalar--tensor theory, i.e., the standard Nambu--Goto coupling to the worldline of a point--like particle or a cosmic brane, the metric to which they couple is $g_{\mu \nu}$ and not $\hat{g}_{\mu \nu}$, and therefore the physical situations described by both cases are no longer equivalent.


Having clarified these issues, we provide examples of different scalar--tensor theories that can be constructed following the prescription described previously.

\subsection{Einstein--AdS Gravity in the Weyl Frame}
\label{sec:wu-eads}


We begin by examining the Weyl uplift of Einstein--AdS gravity, whose action reads

\begin{equation}
I_{\rm{EH}}=\int d^D x \sqrt{-g}\Big(R-2\Lambda \Big) \,. \label{eqn:action-eads}
\end{equation}
Based on the recipe described in the previous section, we consider the scalar--tensor action given by
\begin{equation}
I[\hat{g}_{\mu \nu}]=\int d^D x \sqrt{-\hat{g}}\Big(\hat{R}-2\Lambda \Big) \,, \label{eqn:wu-action-eads}
\end{equation}
where

\begin{equation}
\hat{R}=\frac{1}{2}\delta^{\mu \nu}_{\rho \lambda}\hat{R}^{\rho \lambda}_{\mu \nu} \ , \label{eqn:wu-ricci}
\end{equation}
and $\hat{R}^{\rho \lambda}_{\mu \nu}$ is defined in Eq.~\eqref{eqn:rhat-sourya-tensor}.
Then, from Eqs.~[\eqref{eqn:conf-inv-metric},~\eqref{eqn:rhat-sourya-tensor}--\eqref{eqn:sourya-2nd-trace}], we have that

\begin{equation}
\sqrt{-\hat{g}}=\phi^{D/\Delta}\sqrt{-g} \  \label{eqn:conf-inv-det}
\end{equation}
and

\begin{equation}
\hat{R}=\phi^{-D/\Delta}S=\phi^{-D/\Delta}\left(\phi^{2}R-\frac{4(D-1)}{(D-2)} \phi \Box \phi \right) \,. \label{eqn:rhat-yamabe-eads}
\end{equation}
Thus, we arrive at an expression for the action written as

\begin{equation}
I\left[\hat{g}_{\mu \nu}\right]=I\left[\phi,g_{\mu \nu}\right]=\int d^D x \sqrt{-g}\left[\phi^2 R-\frac{4(D-1)}{(D-2)}\phi \Box \phi - 2\Lambda \phi^{D/\Delta} \right] \,. \label{eqn:wu-action-eads-yamabe}
\end{equation}
Notice that the resulting action is written in terms of a conformally covariant differential operator dubbed the $D$--dimensional Yamabe operator~\cite{Gover:2002ay,Osborn:2015rna}. One can recover the canonical kinetic term of the scalar when integrating by parts, and then the action can be written as

\begin{gather}
I\left[\phi,g_{\mu \nu}\right]=\frac{8(D-1)}{(D-2)}\int \displaylimits_{\mathcal{M}} d^D x \sqrt{-g}\left[\frac{(D-2)}{8(D-1)}\phi^2 R+\frac{1}{2}\partial_{\mu}\phi \partial^{\mu} \phi - \frac{(D-2)}{4(D-1)} \Lambda \phi^{\frac{2D}{D-2}} \right]\nonumber \\
-\frac{4(D-1)}{(D-2)}\int \displaylimits_{\mathcal{\partial M}} d^{D-1}x\sqrt{|h|}n_{\mu}\left(\phi \partial^{\mu}\phi\right) \,. \label{eqn:final-wu-action-eads-yamabe}
\end{gather} 

In this form, one recognizes in the bulk term the standard non--minimally--coupled scalar--tensor theory, which is invariant under Weyl transformation up to a boundary term~\cite{Anastasiou:2022wjq}. The resulting boundary term restores the Weyl symmetry in order to recover the Yamabe operator~\eqref{eqn:wu-action-eads-yamabe}.

In order to derive the EOM of the theory, as discussed in the previous section, one starts with the Einstein's field equations

\begin{equation}
\hat{\epsilon}_{\mu \nu}(\hat{g}) = \hat{R}_{\mu \nu}(\hat{g}) - \frac{1}{2}\hat{R}\hat{g}_{\mu \nu} + \Lambda \hat{g}_{\mu \nu}=0 \,. \label{eqn:wu-eom-eads}
\end{equation}
Then, replacing the conformal invariant metric and using the Weyl transformation laws of the Riemann curvature and its contractions, one obtains

\begin{gather}
\hat{\epsilon}_{\mu \nu}\left(\phi, g_{\mu\nu}\right)= R_{\mu \nu} - \frac{1}{2}Rg_{\mu \nu} + \Lambda \phi^{2/\Delta}g_{\mu \nu} - \frac{2}{\phi}\nabla_\mu \nabla_\nu \phi + \frac{D}{\Delta \phi^2}\nabla_\mu \phi \nabla_\nu \phi \nonumber \\ - \frac{1}{\Delta \phi^2}\nabla_\rho \phi \nabla^\rho \phi  g_{\mu \nu} + \frac{2}{\phi}\Box \phi g_{\mu \nu}=0 \,, \label{eqn:wu-eom-eads4-mai}
\end{gather}
which up to the overall factor $\phi^{2/\Delta}$ recovers the EOM for the scalar--tensor theory as given in Ref.~\cite{Anastasiou:2022wjq}. Taking the trace of the Eq.~\eqref{eqn:wu-eom-eads4-mai}, we obtain

\begin{equation}
 \Box \phi- \frac{D-2}{4\left(D-1\right)} \phi R +\frac{D}{2\left(D-1\right)} \Lambda \phi^{\frac{2}{\Delta}+1} = 0 \,,
\end{equation}
what is the EOM of the scalar field, in complete agreement with Ref.~\cite{Anastasiou:2022wjq}.

Even though this is a rather trivial example of the relation between the Einstein--Hilbert action and the conformally--invariant and non--minimally coupled scalar--tensor theory, it provides the basic principles and necessary tools for the generalization of the prescription for arbitrary diffeomorphism--invariant gravity theories and their corresponding Weyl uplifted scalar--tensor counterparts.

\section{Weyl Uplift of Higher Curvature Gravity Theories}
\label{sec:wuhct}


An important class of theories whose corresponding Weyl--symmetric scalar--tensor versions can be obtained directly with our procedure are higher--curvature theories. The generic case, evaluated on the conformally--invariant metric, can be written as

\begin{equation}
I\left[\hat{g}_{\mu \nu}\right]=\int d^D x \sqrt{-\hat{g}}\mathcal{L}\left(\hat{g}_{\mu \nu},\hat{R}^{\mu}_{~\nu \rho \lambda}\right) \,, \label{eqn:wu-gen-action-hct}
\end{equation}
where the Lagrangian contains all possible contractions between the metric and the Riemann tensor, up to an arbitrary order. Decomposing this action in orders of the Riemann tensor, we can write

\begin{equation}
I\left[\phi,g_{\mu\nu}\right]=\int d^D x \sqrt{-\hat{g}}\left(\lambda_{0,1}\ell^{-2}+\lambda_{1,1}\hat{R}+\sum_{n=2}^{\infty}\sum_{j}\lambda_{n,j}\ell^{2n-2}\hat{\mathcal{R}}_{(n,j)} \right) \,, \label{eqn:wu-gen-action-hct-decomp}
\end{equation}
where $\hat{\mathcal{R}}_{(n,j)}$ is a scalar constructed out of a particular contraction of $n$ Riemann tensors of the form $\hat{R}^{\mu \nu}_ {\rho \lambda}$, and the sum over $j$ is carried over all possible inequivalent contractions.

Considering the relation between $\hat{R}^{\mu \nu}_ {\rho \lambda}$ and $S^{\mu \nu}_{\rho \lambda}$ given in Eq.~\eqref{eqn:rhat-sourya-tensor}, we have that the uplifted scalar--tensor theory obtained from Eq.~\eqref{eqn:wu-gen-action-hct-decomp} can be written as

\begin{gather}
I\left[\phi,g_{\mu\nu}\right]=\int d^D x \sqrt{-g} \Bigg[\lambda_{0,1}\ell^{-2}\phi^{\frac{D}{\Delta
}}+\lambda_{1,1}\left(\phi^2 R -\frac{4(D-1)}{(D-2)}\phi \Box \phi \right)+ \notag \\ \sum_{n=2}^{\infty}\sum_{j}\lambda_{n,j}\ell^{2n-2}\phi^{-(n-1)\frac{D}{\Delta
}}\mathcal{S}_{(n,j)} \Bigg] \,, \label{eqn:wu-genaction-hct-decomp-S}
\end{gather}
where the $\mathcal{S}_{(n,j)}$ corresponds to the same combination of scalars as the ones given by $\hat{\mathcal{R}}_{(n,j)}$, but where the Riemann tensor $\hat{R}^{\mu \nu}_ {\rho \lambda}$ has been replaced by $S^{\mu \nu}_ {\rho \lambda}$. The action derived in Eq.~\eqref{eqn:wu-genaction-hct-decomp-S} defines a whole new class of scalar conformal couplings to higher--curvature theories of gravity and constitutes one of the main results of this work.


In what follows, we provide the derivation of the EOM for these generalized scalar--tensor theories of gravity. Our starting point is the equation of motion of arbitrary higher--curvature gravity, which is given by~\cite{Padmanabhan:2011ex}

\begin{equation}
    E^{\mu}_{\nu} = {P}^{\mu \sigma}_{\rho \lambda}{R}^{\rho \lambda}_{\nu \sigma} - \frac{1}{2}\delta^\mu_\nu {\mathcal{L}} + 2{g}^{\beta \lambda}{\nabla}_\alpha \nabla_\lambda {P}^{\mu \alpha}_{\nu \beta} = 0 \,, \label{eqn:eom-hcg}
\end{equation}
where the tensor $P^{\mu \nu}_{\sigma \rho}$ is defined as

\begin{equation}
 P^{\mu \nu}_{\sigma \rho} \equiv \frac{\partial \mathcal{L}}{\partial R_{\mu \nu}^{\sigma \rho}} \,. \label{eqn:def-P}
\end{equation}
It is convenient to consider the $\mathcal{L}$ as decomposed in terms of the sum of the different scalars constructed from contractions of the Riemann curvatures at different orders, i.e., we consider
\begin{equation}
\mathcal{\hat{L}}=\sum_{n=0}^{\infty}\sum_{j}\lambda_{n,j}\ell^{2n-2}\hat{\mathcal{R}}_{(n,j)} \,, \label{eqn:lag-hcg-decomp}
\end{equation}
where $\hat{\mathcal{R}}_{(0,0)}=1$ and $\hat{\mathcal{R}}_{(0,1)}=\hat{R}$. Then, we can write the P tensor in the Weyl frame, $\hat{P}^{\mu \nu}_{\sigma \rho}$, as

\begin{equation}
\hat{P}^{\mu \nu}_{\sigma \rho} = \sum^{\infty}_{n=1} \sum_{j} \lambda_{n,j}\ell^{2n-2}\phi^{-(n-1)D/\Delta}P^{~~~~~~\mu \nu}_{\phi (n,j)\sigma \rho} \,, \label{eqn:hat-P-hcg}
\end{equation}
where
\begin{equation}
P^{~~~~~~\mu \nu}_{\phi (n,j)\sigma \rho} = \frac{\partial \mathcal{S}_{(n,j)}}{\partial S^{\sigma \rho}_{\mu \nu}} \label{eqn:def-P-phi-S}
\end{equation}
is given by the corresponding derivatives of the $\mathcal{S}_{(n,j)}$ densities with respect to the $S^{\sigma \rho}_{\mu \nu}$ tensor. Note that the $\mathcal{S}_{(n,j)}$ has the same functional form as the $j$--th independent higher curvature scalar of order $n$, but constructed from the $S^{\sigma \rho}_{\mu \nu}$ tensor instead of the Riemann tensor. In total, Eq.~\eqref{eqn:lag-hcg-decomp} obtains the form

\begin{equation}
\hat{\mathcal{L}}=\sum^\infty_{n=0}\sum_{j} \lambda_{n,j}\ell^{2n-2}\phi^{-nD/\Delta}\mathcal{S}_{(n,j)} \,. \label{eqn:wu-lag-hcg}
\end{equation}
Finally, the EOM can be cast in the compact form below:

\begin{equation}
\hat{\epsilon}^\mu_\nu \left(\hat{g}_{\mu \nu}\right)=\hat{E}^{\mu}_{\nu} \left(g_{\mu \nu}, \phi \right) = \hat{P}^{\mu \sigma}_{\rho \lambda}\hat{R}^{\rho \lambda}_{\nu \sigma} - \frac{1}{2}\delta^\mu_\nu \hat{\mathcal{L}} + 2\hat{g}^{\beta \lambda}\hat{\nabla}_\alpha \hat{\nabla}_\lambda \hat{P}^{\mu \alpha}_{\nu \beta} = 0 \,. \label{eqn:wu-eom-hcg}
\end{equation}
Each one of the terms in the previous equation amounts to a combination of the physical metric $g_{\mu \nu}$ and the scalar field $\phi$, given in Eqs.~[\eqref{eqn:rhat-sourya-tensor}, \eqref{eqn:wu-covD}--\eqref{eqn:wu-covar}, \eqref{eqn:hat-P-hcg}--\eqref{eqn:wu-lag-hcg}]. The computation of the uplifted double covariant derivative in the last term of the equation above, is computed in detail in Appendix~\ref{sec:math}.

Furthermore, as shown in Section~\ref{sec:recipe}, the EOM of the scalar field is given by the trace of Eq.~\eqref{eqn:wu-eom-hcg} as

\begin{equation}
\mathcal{E}= \hat{E}^{\mu}_{\mu} \left(g_{\mu \nu}, \phi \right) =\hat{P}^{\mu \sigma}_{\rho \lambda}\hat{R}^{\rho \lambda}_{\mu \sigma} - \frac{D}{2} \hat{\mathcal{L}} + 2\hat{g}^{\beta \lambda}\hat{\nabla}_\alpha \hat{\nabla}_\lambda \hat{P}^{ \alpha}_{\beta} = 0 \,, \label{eqn:wu-eomscalar-hcg}
\end{equation}
where $\hat{P}^{ \alpha}_{\beta}=\hat{P}^{\mu \alpha}_{\mu \beta}$. At this point, we have all the necessary tools to apply the Weyl uplift prescription of higher--curvature gravity theories in some concrete examples.

\subsection{The case of Lovelock Gravity}
\label{subsec:lovelock}
The first example that we will analyze, is the one of Lovelock gravity \cite{Lovelock:1971yv}, which is expressed as a linear combination of Lovelock densities $\mathcal{L}_{k}$ and obtains the general form

\begin{equation}
I\left[g_{\mu \nu}\right]=\int d^D x \sqrt{-g}\left(R-2\Lambda+\sum_{k=2}^{\lfloor\frac{D-1}{2}\rfloor}\lambda_{k}\ell^{2k-2}\mathcal{L}_{k} \right) \,, \label{eqn:wu-action-lovelock}
\end{equation}
where 

\begin{equation}
\mathcal{L}_{k} = \frac{1}{2^k}\delta^{\mu_1 ...\mu_{2k}}_{\nu_1 ...\nu_{2k}}R^{\nu_1 \nu_2}_{\mu_1 \mu_2}...R^{\nu_{2k-1}\nu_{2k}}_{\mu_{2k-1}\mu_{2k}} \,. \label{eqn:wu-defn-density-lovelock}
\end{equation}
Following the prescription that we have described previously, by replacing all the geometric objects by the hatted ones, we generate the Weyl uplifted version of Lovelock theory, using Eq.~\eqref{eqn:wu-genaction-hct-decomp-S}. Then, the corresponding scalar--tensor theory constructed out of conformal coupling of scalar fields to Lovelock densities becomes

\begin{gather}
I[g_{\mu \nu}, \phi]=\int d^D x \sqrt{-g}\Bigg[-2\Lambda \phi^{D/\Delta} + \left(\phi^2 R -\frac{4(D-1)}{(D-2)}\phi \Box \phi \right) + \nonumber \\ \sum_{k=2}^{\lfloor\frac{D-1}{2}\rfloor}\frac{\lambda_{k}\ell^{2k-2}\phi^{-(k-1)D/\Delta}}{2^k}\delta_{\nu_1 ...\nu_{2k}}^{\mu_1 ...\mu_{2k}}S_{\mu_1 \mu_2}^{\nu_1 \nu_2}...S_{\mu_{2k-1}\mu_{2k}}^{\nu_{2k-1}\nu_{2k}}\Bigg] \,, \label{eqn:wu-action-lovelock-sourya}
\end{gather}
in agreement with Ref.~\cite{Oliva:2011np}. The EOM of this theory can be obtained directly from Eq.~\eqref{eqn:wu-eom-hcg}, which corresponds to the Lanczos--Lovelock tensor evaluated on the conformally invariant metric, that reads


\begin{equation}
\hat{E}^{\mu}_{\nu} =\hat{R}^{\mu}_{\nu} -\frac{1}{2}(\hat{R}-2\Lambda)\delta^{\mu}_{\nu}-\hat{H}^{\mu}_{\nu}=0 \ , \label{eqn:wu-eom-lovelock}
\end{equation}
where

\begin{equation}
\hat{H}^{\mu}_{\nu}=\sum_{k=2}^{{\lfloor\frac{D-1}{2}\rfloor}}\frac{\lambda_{k}\ell^{2k-2}}{2^{(k+1)}}\delta^{\mu \mu_1 ...\mu_{2k}}_{\nu \nu_1 ...\nu_{2k}}\hat{R}^{\nu_1 \nu_2}_{\mu_1 \mu_2}...\hat{R}^{\nu_{2k-1}\nu_{2k}}_{\mu_{2k-1}\mu_{2k}} \,. \label{eqn:defn-LL}
\end{equation}
Then, expressing all the hatted quantities in terms of the fields defined in the Weyl frame, we obtain the EOM of the metric as

\begin{gather}
\hat{E}_{\nu}^{\mu} \left(g_{\mu \nu},\phi \right)=R_{\nu}^{\mu} - \frac{1}{2}R\delta_{\nu}^{\mu} + \Lambda \phi^{2/\Delta}\delta_{\nu}^{\mu} - \frac{2}{\phi}\nabla^\mu \nabla_\nu \phi + \frac{D}{\Delta \phi^2}\nabla^\mu \phi \nabla_\nu \phi \nonumber \\ - \frac{1}{\Delta \phi^2}\nabla^\rho \phi \nabla_\rho \phi  \delta_{\nu}^{\mu} + \frac{2}{\phi}\Box \phi \delta_{\nu}^{\mu}-\phi^{2/\Delta}\mathcal{H}_{\nu}^{\mu}=0 \,, \label{eqn:final-wu-eom-lovelock}
\end{gather} where

\begin{equation}
\mathcal{H}_{\nu}^{\mu}=\sum_{k=2}^{{\lfloor\frac{D-1}{2}\rfloor}}\frac{\lambda_{k}\ell^{2k-2}\phi^{-kD/\Delta}}{2^{(k+1)}}\delta^{\mu \mu_1 ...\mu_{2k}}_{\nu \nu_1 ...\nu_{2k}}S^{\nu_1 \nu_2}_{\mu_1 \mu_2}...S^{\nu_{2k-1}\nu_{2k}}_{\mu_{2k-1}\mu_{2k}} \label{eqn:wu-LL}
\end{equation}
is the Weyl uplift of the Lanczos--Lovelock tensor. Then, since the derivation of the EOM of the scalar field amounts to the computation of the trace of Eq.~\eqref{eqn:final-wu-eom-lovelock}, we get

\begin{gather}
\mathcal{E}= \phi \Box \phi- \frac{D-2}{4\left(D-1\right)} \phi R +\frac{D (D-2)}{4\left(D-1\right)} \phi^{\frac{2}{\Delta}+1} \notag \\ -\frac{\phi^{\frac{2}{\Delta}+1}}{D-1}\sum_{k=2}^{{\lfloor\frac{D-1}{2}\rfloor}}\frac{\lambda_{k}\ell^{2k-2}\left(D-2k\right)}{2^{(k+2)} }\phi^{-kD/\Delta}\delta^{\mu_1 ...\mu_{2k}}_{\nu_1 ...\nu_{2k}}S^{\nu_1 \nu_2}_{\mu_1 \mu_2}...S^{\nu_{2k-1}\nu_{2k}}_{\mu_{2k-1}\mu_{2k}} =0 \,.
\label{eomscalarLovelock}
\end{gather}
Both Eqs.~\eqref{eqn:final-wu-eom-lovelock} and~\eqref{eomscalarLovelock} are in agreement with the field equations derived in Ref.~\cite{Oliva:2011np}.

\subsection{The case of Quasitopological and Generalized Quasitopological Gravity}
\label{subsec:qtg}

Lovelock theories of gravity belong to a more generic class of theories, dubbed quasitopological gravities~\cite{Oliva:2010eb,Myers:2010ru,Myers:2010jv,Cisterna:2017umf,Dehghani:2013ldu}. Even though the former has second--order equations of motion whereas the EOM of the latter are of fourth--order, they share similar features when evaluated on static and spherically symmetric backgrounds~\cite{Bueno:2019ycr}. In particular, the trace of their EOM is proportional to the Lagrangian of the theory, a property that is satisfied by Lovelock theories as well. An interesting sub--class of quasitopological gravities has been constructed in odd bulk dimension and is given in Ref.~\cite{Oliva:2010zd}.

By analogy to the Lovelock class of theories, we provide the Weyl uplift of the quasitopological densities of the particular type mentioned above, by evaluating the corresponding action in the conformally invariant metric. Therefore, for the $k$--th order in the curvature densities we have

\begin{eqnarray}
I_{k} \left(\hat{g}_{\mu \nu}\right)=\int d^D x \sqrt{-\hat{g}}\Bigg[\frac{1}{2^{k} \left(D-2k+1\right) } \delta^{\mu_1 ...\mu_{2k}}_{\nu_1 ...\nu_{2k}} \left(\hat{W}^{\nu_1 \nu_2}_{\mu_1 \mu_2}...\hat{W}^{\nu_{2k-1}\nu_{2k}}_{\mu_{2k-1}\mu_{2k}} - \hat{R}^{\nu_1 \nu_2}_{\mu_1 \mu_2}...\hat{R}^{\nu_{2k-1}\nu_{2k}}_{\mu_{2k-1}\mu_{2k}}\right) \notag \\ 
&&\hskip -10cm - \ c_{k} \hat{W}^{\mu_{2k-1} \mu_{2k}}_{\mu_1 \mu_2} \hat{W}^{\mu_1 \mu_2}_{\mu_3 \mu_4}...\hat{W}^{\mu_{2k-3}\mu_{2k-2}}_{\mu_{2k-1}\mu_{2k}}\Bigg] \,,
\label{hatquasitopological}
\end{eqnarray}
where


\begin{equation}
c_{k} = \frac{\left(D-4\right)!}{\left(D-2k+1\right)!}\frac{\left[k\left(k-2\right)D\left(D-3\right)+k\left(k+1\right)\left(D-3\right)+\left(D-2k\right)\left(D-2k-1\right)\right]}{\left[\left(D-3\right)^{k-1} \left(D-2\right)^{k-1}+2^{k-1}-2\left(3-D\right)^{k-1}\right]}\,.
\label{quasitop_coeff}
\end{equation}
These densities are defined in dimension $D=2k-1$  for arbitrary values of $k$. Taking into account the transformation of the Weyl tensor as $\hat{W}^{\alpha \beta}_{\mu \nu}=\phi^{-2/\Delta} W^{\alpha \beta}_{\mu \nu}$ along with Eq.~\eqref{eqn:rhat-sourya-tensor}, we obtain

\begin{eqnarray}
I_{k} \left(g_{\mu \nu},\phi \right)=\int d^D x \sqrt{-g} \phi^{\left(D-2k\right)/\Delta} \Bigg[\frac{1}{2^{k} \left(D-2k+1\right)} \delta^{\mu_1 ...\mu_{2k}}_{\nu_1 ...\nu_{2k}} \Big(W^{\nu_1 \nu_2}_{\mu_1 \mu_2}...W^{\nu_{2k-1}\nu_{2k}}_{\mu_{2k-1}\mu_{2k}} - \notag \\  &&\hskip -12cm \phi^{-k\left(D-2\right)/\Delta} S^{\nu_1 \nu_2}_{\mu_1 \mu_2}...S^{\nu_{2k-1}\nu_{2k}}_{\mu_{2k-1}\mu_{2k}}\Big)
 - c_{k} W^{\mu_{2k-1} \mu_{2k}}_{\mu_1 \mu_2} W^{\mu_1 \mu_2}_{\mu_3 \mu_4}...W^{\mu_{2k-3}\mu_{2k-2}}_{\mu_{2k-1}\mu_{2k}}\Bigg] \,,
\label{weylquasitopological}
\end{eqnarray} 
which expresses the conformal coupling of a scalar field to the quasitopological densities here considered.

The aforementioned class of theories belong to a bigger family of gravity theories, called Generalized quasitopological gravities (GQTs). The latter are characterised by the presence of spherically symmetric solutions whose lapse function satisfies a defining relation which can be a differential equation of up to second order in radial derivatives. However, unlike the quasitopological densities, the corresponding equation for the lapse is not algebraic~\cite{Bueno:2022res}. The simplest example of this class of theories can be found in 4D, with Einsteinian Cubic Gravity (ECG)~\cite{Bueno:2016xff}, whose action is given by

\begin{equation}
I = \int d^4 x \sqrt{-g} \Big(\frac{1}{2\kappa}[R - 2\Lambda] + \kappa \lambda \mathcal{P}\Big) \ \ , \ \ \kappa = 8\pi G \,, 
\label{eqn:ecg-action}
\end{equation}
where
\begin{equation}
\mathcal{P} = 12 R^{~\rho ~\sigma}_{\mu ~\nu} R^{~\lambda ~\delta}_{\rho ~\sigma} R^{~\mu ~\nu}_{\lambda ~\delta} + R^{~~~\mu \nu}_{\rho \sigma}R^{~~~\rho \sigma}_{\lambda \delta} R^{~~~\lambda \delta}_{\mu \nu} - 12 R_{\mu \nu \rho \sigma} R^{\mu \rho} R^{\nu \sigma} + 8 R^{~\nu}_{\mu} R^{~\rho}_{\nu} R^{~\mu}_{\rho} \,,
\label{eqn:ecg-cubic-term}
\end{equation}
and $\lambda$ is a dimensionless coupling constant. The EOM of this theory can be obtained from Eq.~\eqref{eqn:eom-hcg}, considering that $\mathcal{L}$ is the Lagrangian density given by the parenthesis of Eq.~\eqref{eqn:ecg-action} and the $P$--tensor is given by~\cite{Hennigar:2016gkm}
\begin{eqnarray}
&&\hskip -13.3cm P^{cd}_{ab} = \frac{\partial \mathcal{L}}{\partial R^{ab}_{cd}} \nonumber \\ = \frac{1}{4\kappa}\delta^{cd}_{ab} + 12 \kappa \lambda \Big(R^{[d}_a R^{c]}_b + 2\delta^{[d}_{[b} R^{|m|}_{a]} R^{c]}_m + 2\delta^{[c}_{[b} R^{~~~d]}_{a]m ~n} R^{mn} + 3 R^{~m[c|n|}_{a} R^{~~~d]}_{bm ~n} \nonumber \\ + \frac{1}{4} R^{mn}_{ab}R^{cd}_{mn}\Big) \,.
\label{eqn:ecg-p-tensor}
\end{eqnarray}
Using Eq.~\eqref{eqn:ecg-p-tensor}, one can find the explicit form of the EOM, which we present in Appendix~\ref{sec:ecg-eom-explicit}. Then, following our prescription, it is straightforward to derive the scalar--tensor theory by uplifting the corresponding action in the Weyl frame, which takes the form
\begin{eqnarray}
\hat{I}[g_{\mu \nu}, \phi]=\int d^4 x \sqrt{-g}\phi^4 \hat{\mathcal{L}}  \,,
\label{ecgweyl}
\end{eqnarray}
where the uplifted Lagrangian density is given by
\begin{eqnarray}
    \hat{\mathcal{L}} = \Bigg[\frac{1}{2\kappa}\Big(-2\Lambda  + \phi^{-4}\left[\phi^2 R -6\phi \Box \phi \right]\Big) + \kappa \lambda \phi^{-12} \Big(12 S^{~\sigma ~\delta}_{\alpha ~\beta} S^{~\mu ~\nu}_{\sigma ~\delta} S^{~\alpha ~\beta}_{\mu ~\nu} + \notag \\ 
&&\hskip -10cm  S^{~~~\sigma \delta}_{\alpha \beta} S^{~~~\mu \nu}_{\sigma \delta} S^{~~~\alpha \beta}_{\mu \nu} - 12 S_{\alpha \beta \sigma \delta} S^{\alpha \sigma} S^{\beta \delta} + 8 S^{~\beta}_{\alpha} S^{~\sigma}_{\beta} S^{~\alpha}_{\sigma}\Big)\Bigg] \,,
\end{eqnarray}
and the $S$--tensor is defined in Eq.~\eqref{eqn:sourya}. Then, the EOM for the metric field in the Weyl frame becomes
\begin{equation}
\hat{\epsilon}^\mu_\nu \left(\hat{g}_{\mu \nu}\right) = \hat{P}^{\mu \sigma}_{\rho \lambda}\hat{R}^{\rho \lambda}_{\nu \sigma} - \frac{1}{2}\delta^\mu_\nu \hat{\mathcal{L}} + 2\hat{g}^{\beta \lambda}\hat{\nabla}_\alpha \hat{\nabla}_\lambda \hat{P}^{\mu \alpha}_{\nu \beta} = 0 \,, \label{eqn:wu-ecg-eom}
\end{equation}
where
\begin{eqnarray}
\hat{P}^{cd}_{ab} = \frac{1}{4\kappa}\delta^{cd}_{ab} + 12 \kappa \lambda \phi^{-8}\Big(S^{[d}_a S^{c]}_b + 2\delta^{[d}_{[b} S^{|m|}_{a]} S^{c]}_m + 2\delta^{[c}_{[b} S^{~~~d]}_{a]m ~n} S^{mn} + \nonumber \\ 
&&\hskip -5cm 3 S^{~m[c|n|}_{a} S^{~~~d]}_{bm ~n} + \frac{1}{4} S^{mn}_{ab}S^{cd}_{mn}\Big) \,,
\end{eqnarray}
and
\begin{equation}
    \hat{g}^{\mu \nu}=\phi^{-2}g^{\mu \nu} \ \ , \ \ \hat{R}_{\lambda \rho}^{\mu \nu}=\phi^{-4}S^{\mu \nu}_{\lambda \rho} \,,
    \label{eqn:4d-ecg-trans-law}
\end{equation}
with the $\hat{\nabla}_\alpha \hat{\nabla}_\lambda \hat{P}^{\mu \alpha}_{\nu \beta}$ term given in Appendix~\ref{sec:math}. Moreover, as mentioned earlier, one can take the trace of the EOM given in Eq.~\eqref{eqn:ecg-eom-explicit}, and then Weyl uplift it to get the EOM for the scalar field $\phi$, i.e., $\hat{\epsilon}=\hat{\epsilon}^\mu_\mu =0$.

The examples presented above confirm that the prescription of Weyl uplift of metric theories is a powerful tool for the determination of new classes of conformally coupled scalar--tensor theories.

\section{Weyl Uplift and Renormalization}
\label{sec:wu-ren}


Recent works have pointed out the significance of bulk Weyl invariance of metric theories in the cancellation of IR divergences~\cite{Grumiller:2013mxa,Anastasiou:2016jix,Anastasiou:2020mik}. The applicability of this new principle for the renormalization of gravitational actions has been extended to a particular class of scalar--tensor theories~\cite{Barrientos:2022yoz,Anastasiou:2022wjq}, that being the bulk part of the action in Eq.~\eqref{eqn:final-wu-action-eads-yamabe} in $D=4$. The restoration of the Weyl invariance of the action of the theory, such that it is fully symmetric --- and not up to a boundary term --- leads to the Yamabe action of Eq.~\eqref{eqn:wu-action-eads-yamabe}, or equivalently Eq.~\eqref{eqn:final-wu-action-eads-yamabe}. This action is partially renormalized, since there are certain configurations in the solutions space (constant scalar field), where infinities still arise. Indeed, the $\phi=1$ choice in the action of Eq.~\eqref{eqn:wu-action-eads-yamabe} leads to Einstein--AdS gravity as given in Eq.~\eqref{eqn:action-eads}, which is both divergent and breaks the Weyl invariance of the action. 

In order to remedy this problem, our starting point will be the Einstein--AdS action enhanced by the Gauss--Bonnet density with a fixed coupling constant

\begin{equation}\label{LagEHGB}
I_{ren} =\frac{1}{16 \pi G_N} \int \displaylimits_{\mathcal{M}} d^4 x \sqrt{|g|} \left(R+\frac{6}{\ell^2} + \frac{\ell^2}{4} E_{4}\right) \,,
\end{equation}
where
\begin{equation}
E_{4} = \dfrac{1}{4} \delta _{\nu _{1} \ldots  \nu _{4}}^{\mu _{1} \ldots  \mu _{4}} R_{\mu _{1} \mu _{2}}^{\nu _{1} \nu _{2}} R_{\mu _{3} \mu _{4}}^{\nu _{3} \nu _{4}} = R^2 - 4R^\mu_\nu R^\nu_\mu + R^{\mu\nu}_{\lambda\rho}R^{\lambda\rho}_{\mu\nu} \,.
\end{equation}
The latter corresponds to the topologically renormalized Einstein--AdS action, which is equivalent to the action obtained considering the standard holographic renormalization counterterms for generic asymptotically locally AdS (AlAdS) manifolds in 4D~\cite{Miskovic:2009bm}. The latter can be written in a MacDowell--Mansouri form~\cite{MacDowell:1977jt,Miskovic:2009bm} as\footnote{For completeness, one should consider the additional Euler characteristic contribution accompanying the above action, however we omit it for simplicity.} 

\begin{equation}
I_{\rm{MM}}[g_{\mu \nu}]=\frac{\ell^2}{4}\int \displaylimits_{\mathcal{M}} d^4 x \sqrt{-g}\delta^{\mu_1 \mu_2 \mu_3 \mu_4}_{\nu_1 \nu_2 \nu_3 \nu_4}\left(R_{\mu_1 \mu_2}^{\nu_1 \nu_2 }+\frac{1}{\ell^2}\delta_{\mu_1 \mu_2}^{\nu_1 \nu_2 }\right)\left(R_{\mu_3 \mu_4}^{\nu_3 \nu_4 }+\frac{1}{\ell^2}\delta_{\mu_3 \mu_4}^{\nu_3 \nu_4 }\right) \,. \label{eqn:ren-action-eads-mm}
\end{equation}
The field equations are still of second order since the initial action has been enhanced by a topological invariant. Of course, for $D \neq 4$, the Gauss--Bonnet density is no longer topological and the corresponding EOM are of fourth order\footnote{In $D$ dimensions this action is divergent and additional counterterms have to be added in order to render it finite.}. Then the uplifted version of this action in the Weyl frame in arbitrary $D$ dimensions obtains the form

\begin{equation}
I_{\rm{MM}}[\phi, g_{\mu \nu}]=\frac{\ell^2}{4}\int \displaylimits_{\mathcal{M}} d^D x \sqrt{-g}\phi^{\frac{D}{\Delta}}\delta^{\mu_1 \mu_2 \mu_3 \mu_4}_{\nu_1 \nu_2 \nu_3 \nu_4}\left(\phi^{-\frac{D}{\Delta}}S_{\mu_1 \mu_2}^{\nu_1 \nu_2 }+\frac{1}{\ell^2}\delta_{\mu_1 \mu_2}^{\nu_1 \nu_2 }\right)\left(\phi^{-\frac{D}{\Delta}}S_{\mu_3 \mu_4}^{\nu_3 \nu_4 }+\frac{1}{\ell^2}\delta_{\mu_3 \mu_4}^{\nu_3 \nu_4 }\right) \,. \label{eqn:wu-ren-action-eads-mm-sourya}
\end{equation}
Following the notation of Ref.~\cite{Anastasiou:2022wjq}, we define
\begin{equation}
\Sigma_{\rho \lambda}^{\mu \nu} =\frac{1}{\phi^2}\left(S_{\rho \lambda}^{\mu \nu}+\frac{\phi^{D/\Delta}}{\ell^2} \delta_{\rho \lambda}^{\mu \nu} \right) \,, \label{eqn:def-sigma}
\end{equation}
and therefore we can write the action as
\begin{equation}
I_{\rm{MM}}[\phi, g_{\mu \nu}]=\frac{\ell^2}{4}\int \displaylimits_{\mathcal{M}} d^D x \sqrt{-g}\phi^{\left(4-\frac{D}{\Delta}\right)}\delta^{\mu_1 \mu_2 \mu_3 \mu_4}_{\nu_1 \nu_2 \nu_3 \nu_4}\Sigma_{\mu_1 \mu_2}^{\nu_1 \nu_2 }\Sigma_{\mu_3 \mu_4}^{\nu_3 \nu_4 } \,. \label{eqn:wu-ren-action-eads-mm-sigma}
\end{equation}
In the case of $D=4$, the action reduces to
\begin{equation}
I_{\rm{MM}}[\phi, g_{\mu \nu}]=\frac{\ell^2}{4}\int \displaylimits_{\mathcal{M}} d^4 x \sqrt{-g}\delta^{\mu_1 \mu_2 \mu_3 \mu_4}_{\nu_1 \nu_2 \nu_3 \nu_4}\Sigma_{\mu_1 \mu_2}^{\nu_1 \nu_2 }\Sigma_{\mu_3 \mu_4}^{\nu_3 \nu_4 } \ , \label{eqn:wu-ren-action-eads4-mm-sigma}
\end{equation}
which is the renormalized scalar--tensor action with Weyl symmetry considered in~\cite{Anastasiou:2022wjq}.

The action of Eq.~\eqref{eqn:wu-ren-action-eads4-mm-sigma} defines the same theory as Eq.~\eqref{eqn:wu-action-eads-yamabe} when $D=4$, as it leads to the same EOM. However, there is an interesting issue regarding the relation between the two actions in the context of renormalization by bulk local scale invariance, previously discussed in Refs.~\cite{Grumiller:2013mxa,Anastasiou:2016jix,Anastasiou:2020mik}. In particular, both actions differ by topological and boundary terms, which ensure that Eq.~\eqref{eqn:wu-ren-action-eads4-mm-sigma} is finite for arbitrary field configurations that are solutions of the theory. As explained in Ref.~\cite{Anastasiou:2022wjq}, the action of Eq.~\eqref{eqn:wu-action-eads-yamabe}, is divergent for field configurations of constant $\phi$ in AlAdS spaces, where the divergence is proportional to the volume of the manifold. These configurations can be understood as singular points of the Weyl symmetry, in the sense that for constant $\phi$, the local scale symmetry is broken. On the other hand, the action of Eq.~\eqref{eqn:wu-ren-action-eads4-mm-sigma} is both finite and Weyl invariant even for these solutions, which provides more evidence on the relation between bulk Weyl symmetry and renormalization.

Furthermore, the difference between the actions of Eqs.~\eqref{eqn:wu-ren-action-eads4-mm-sigma} and \eqref{eqn:wu-action-eads-yamabe} can be traced back to the parent actions, defined in the Einstein frame. In particular, Eqs.~\eqref{eqn:action-eads} and~\eqref{eqn:ren-action-eads-mm} are equivalent up to the addition of the 4D Euler density that cancels the divergences and renders the Einstein--AdS action finite. This relation suggests that only the Weyl uplift from renormalized gravitational actions provides scalar--tensor theories of gravity which are finite for all possible configurations of the fields of the theory.


\section{Summary}
\label{sec:discuss}


In this work we have developed a method for constructing Weyl invariant scalar--tensor theories starting from purely metric theories, as discussed in section~\ref{sec:recipe}. The procedure consists in replacing the metric by a rank--(0,2) tensor, dubbed the conformal invariant metric and defined as $\phi^{\frac{2}{\Delta}}g_{\mu \nu}$, and then evaluating the Riemannian curvature and covariant derivative operator appearing in the original action on this new object instead of the metric. 

Through this algorithm, we manage to re--derive the Weyl--invariant scalar--tensor theory of Eq.~\eqref{eqn:wu-gen-action} and its renormalized version of Eq.~\eqref{eqn:wu-ren-action-eads-mm-sigma}, both of which have the same equations of motion and contain Einstein--AdS gravity as part of its solution space (see section~\ref{sec:wu-eads}). We provide the generic Lagrangian that describes the conformal scalar couplings to an arbitrary higher-curvature gravity theory. The resulting formula successfully reproduced the scalar conformal couplings to Lovelock gravity, which were introduced in Ref.~\cite{Oliva:2011np}, as discussed in section~\ref{subsec:lovelock}. Furthermore, we derive the 
field equation of the metric and scalar fields. In particular, the EOM of the metric fields can be determined by expressing the original field equations in the Weyl frame. Interestingly enough, we show that the field equations of the scalar field are determined by the trace of the EOM of the metric. This result is in agreement with the tracelessness of the energy-momentum tensor in the presence of conformal symmetry.
A concrete application of this prescription, has been presented in section~\ref{subsec:qtg}, where we provide the scalar conformal couplings of quasitopological gravity and Einsteinian cubic gravity.

It becomes evident from our analysis, that the renormalization of this class of scalar--tensor theories passes through the requirement of exhibiting Weyl invariance for all configurations allowed by the field equations. This feature becomes manifest in the 4D conformally coupled scalar theory to the Einstein--Hilbert action. A formal proof of this principle to higher dimensions and higher--curvature theories of gravity will follow in future work.

In forthcoming research, it would be possible to implement the renormalization of higher--dimensional Einstein--AdS gravity and of Lovelock gravity by restoring on--shell Weyl invariance, and then to construct the conformal scalar couplings to these theories. This could be done in analogy to the Einstein--AdS case in four dimensions presented in section~\ref{sec:wu-ren}.

\section*{Acknowledgements}
The work of GA is funded by ANID, Convocatoria Nacional Subvenci\'on a Instalaci\'on en la Academia Convocatoria A\~no 2021, Folio SA77210007. The work of IJA is funded by ANID FONDECYT grants No.~11230419 and~1231133, and by ANID Becas Chile grant No.~74220042. IJA also acknowledges funding by ANID, REC Convocatoria Nacional Subvenci\'on a Instalaci\'on en la Academia Convocatoria A\~no 2020, Folio PAI77200097. IJA is grateful to M\'aximo Ba\~nados for insightful discussions and to Andrei Parnachev and the School of Mathematics at Trinity College Dublin for their hospitality. The work of AC is funded by ANID/ACT210100 and ANID FONDECYT grant No.~3230222. AC would like to express sincere gratitude to the organizers of the NordGrav@ICEN 2023 for hosting an incredibly stimulating event. AC also deeply appreciates the warm hospitality extended by IJA and ICEN, UNAP, Iquique, where a significant portion of the work was carried out.

\appendix

\section{Equation of Motion for ECG Theory}
\label{sec:ecg-eom-explicit}
In section~\ref{subsec:qtg} we presented the Weyl uplift of ECG. Before the uplift, the explicit form of the EOM is given by 
\begin{align}
\epsilon^i_j =& \frac{1}{4\kappa}(R + 2\Lambda)\delta^i_j - 4\delta^i_j R^c_a R^{ab} R_{bc} + 12R_{ab} R^{ia} R^b_j + 6\delta^i_j R^{ab} R^{cd} R_{acbd} - 12R^{ab} R_{acbd} R^{ic ~d}_{~~j} \nonumber \\ & -  \frac{1}{2} \delta^i_j R^{mn}_{ab} R^{abcd} R_{cdmn} - 12 R^{bc} R^a_j R^i_{~bac} + 12 R^c_a R^{ab} R^i_{~bjc} - 6\delta^i_j R^{~m ~n}_{a ~c} R^{abcd} R_{bmdn} +  \nonumber \\ &    3R_{bcdm} R^{iabc} R^{dm}_{ja} -  12 R^{bc} R^{ia} R_{jbac} + 36R_{adcm} R^{iabc} R^{~d ~m}_{j ~b} - 12R^{ia} \nabla_a \nabla_b R^b_j - \nonumber \\ &   6R^{~abc}_j \nabla_a \nabla_d R^{id}_{bc} + 24\nabla^a R^i_j \nabla_b R^b_a -  12\nabla_a R^{ia} \nabla_b R^b_j + 12R^i_j \nabla_b \nabla_a R^{ab} - 12 R^a_j \nabla_b \nabla_a R^{ib} + \nonumber \\ &     12R^{ab} \nabla_b \nabla_a R^i_j - 12R^a_j \nabla_b \nabla^b R^i_a - 12R^{ia} \nabla_b \nabla^b R_{ja} - 12\delta^i_j R^{ab} \nabla_b \nabla_c R^c_a + 12 R^i_a \nabla_b \nabla_j R^b_a + \nonumber \\ &  12R^{ab} \nabla_b \nabla_j R^i_a - 12\nabla_a R_{jb} \nabla^b R^{ia} -  24\nabla_b R_{ja} \nabla^b R^{ia} - 12\delta^i_j \nabla_a R^{ab} \nabla_c R^c_b - 12\delta^i_j R^{ab} \nabla_c \nabla_b R^c_a  \nonumber \\ & + 12R^i_{~ajb} \nabla_c \nabla^c R^{ab} + 12 R^{ab} \nabla_c \nabla^c R^i_{~ajb} +  36R^{iabc} \nabla_c \nabla_d R^{~~~d}_{jba} + 12R^{ab} \nabla_c \nabla_j R^i_{~abc} - \nonumber \\ & 12\delta^i_j \nabla_b R_{ac} \nabla^c R^{ab} + 24\nabla_c R^i_{~ajb} \nabla^c R^{ab} +  12R^i_{~abc} \nabla^c \nabla_j R^{ab} +  24\delta^i_j \nabla^c R^{ab} \nabla_d R_{acbd} - \nonumber \\ & 36\nabla^c R^{ia ~b}_{~~j} \nabla_d R^{~~~d}_{acb} - 36\nabla^c R^{ia ~b}_{~~j} \nabla_d R^{~d}_{a ~bc} -  36\nabla_b R^{iabc} \nabla_d R_{jcad} - 6\nabla_a R^{iabc} \nabla_d R^{~d}_{j ~bc} - \nonumber \\ & 6R^{iabc} \nabla_d \nabla_a R^{~d}_{j ~bc} - 36R^{abcd} \nabla_d \nabla_b R^i_{ajc} +  12\delta^i_j R^{ab} \nabla_d \nabla_c R^{~c ~d}_{a ~b} - 36R^{ia ~b}_{~~j} \nabla_d \nabla_c R^{~c ~d}_{a ~b} + \nonumber \\ & 36R^{~abc}_j \nabla_d \nabla_c R^{i ~~d}_{~ba} - 6\nabla_a R_{jdbc} \nabla^d R^{iabc} +  36\nabla_c R_{jbad} \nabla^d R^{iabc} + 12\delta^i_j R_{acbd} \nabla^d \nabla^c R^{ab} + \nonumber \\ & 12\nabla^b R^a_j \nabla^i R_{ab} + 12\nabla_c R_{jabc} \nabla^i R^{ab} +  12\nabla_b R^b_a \nabla^i R^a_j +  12\nabla^c R^{ab} \nabla^i R_{jabc} + \nonumber \\ & 12R^a_j \nabla^i \nabla_b R^b_a +  12R^{ab} \nabla^i \nabla_b R_{ja} + 12R^{ab}\nabla^i \nabla_c R^{~~~c}_{jab} + 12R_{jabc} \nabla^i \nabla^c R^{ab} + \nonumber \\ & 12\nabla^b R^{ia} \nabla_j R_{ab} +  12\nabla_c R^i_{~abc} \nabla_j R^{ab} + 12\nabla_b R^b_a \nabla_j R^{ia} + 12\nabla^c R^{ab} \nabla_j R^i_{~abc} \,. 
\label{eqn:ecg-eom-explicit}
\end{align}

\section{Computation of the Double Uplifted Covariant Derivative Term}
\label{sec:math}

In the computation of the EOM for the Weyl uplift of Higher Curvature Gravity, which is given in Eq.~\eqref{eqn:wu-eom-hcg}, the term $\hat{\nabla}_\alpha \hat{\nabla}_\lambda \hat{P}^{\mu \alpha}_{\nu \beta}$ has to be computed. We do so in this section. 

By considering the uplift of the covariant derivative, as defined in Eqs.~\eqref{eqn:wu-covD}--\eqref{eqn:wu-covar}, we have that
\begin{equation}
\hat{\nabla}_\lambda \hat{P}^{\mu \alpha}_{\nu \beta} = \nabla_\lambda \hat{P}^{\mu \alpha}_{\nu \beta} + \underbrace{\Theta^\mu_{\lambda \sigma} \hat{P}^{\sigma \alpha}_{\nu \beta} + \Theta^\alpha_{\lambda \sigma} \hat{P}^{\mu \sigma}_{\nu \beta} - \Theta^\sigma_{\lambda \nu} \hat{P}^{\mu \alpha}_{\sigma \beta} - \Theta^\sigma_{\lambda \beta} \hat{P}^{\mu \alpha}_{\nu \sigma}}_{A^{\mu \alpha}_{\lambda \nu \beta}} \,. \label{eqn:wu-single-covD}
\end{equation}
Then, by defining a new tensor $A^{\mu \alpha}_{\lambda \nu \beta}$ in terms of the parts of Eq.~\eqref{eqn:wu-single-covD} that depend on the modification to the connection $\Theta^\sigma_{\lambda \beta}$, we can compute the second uplifted covariant derivative as
\begin{equation}
\hat{\nabla}_\alpha \hat{\nabla}_\lambda \hat{P}^{\mu \alpha}_{\nu \beta} = \nabla_\alpha \nabla_\lambda \hat{P}^{\mu \alpha}_{\nu \beta} + \Theta^\mu_{\alpha \sigma} A^{\sigma \alpha}_{\lambda \nu \beta} + \Theta^\alpha_{\alpha \sigma} A^{\mu \sigma}_{\lambda \nu \beta} - \Theta^\sigma_{\alpha \lambda} A^{\mu \alpha}_{\sigma \nu \beta} - \Theta^\sigma_{\alpha \nu} A^{\mu \alpha}_{\lambda \sigma \beta} - \Theta^\sigma_{\alpha \beta} A^{\mu \alpha}_{\lambda \nu \sigma} \,. \label{eqn:wu-double-covD}
\end{equation}
Finally, replacing the definition of $A^{\mu \sigma}_{\lambda \nu \beta}$, we obtain
\begin{gather}
\hat{\nabla}_\alpha \hat{\nabla}_\lambda \hat{P}^{\mu \alpha}_{\nu \beta} = \nabla_\alpha \nabla_\lambda \hat{P}^{\mu \alpha}_{\nu \beta} + \Theta^\mu_{\alpha \sigma} \Theta^\sigma_{\lambda \rho} \hat{P}^{\rho \alpha}_{\nu \beta} + \Theta^\mu_{\alpha \sigma} \Theta^\alpha_{\lambda \rho} \hat{P}^{\sigma \rho}_{\nu \beta} - \Theta^\mu_{\alpha \sigma} \Theta^\rho_{\lambda \nu} \hat{P}^{\sigma \alpha}_{\rho \beta} - \Theta^\mu_{\alpha \sigma} \Theta^\rho_{\lambda \beta} \hat{P}^{\sigma \alpha}_{\nu \rho} \nonumber \\  + \Theta^\alpha_{\alpha \sigma} \Theta^\mu_{\lambda \rho} \hat{P}^{\rho \sigma}_{\nu \beta} + \Theta^\alpha_{\alpha \sigma} \Theta^\sigma_{\lambda \rho} \hat{P}^{\mu \rho}_{\nu \beta} - \Theta^\alpha_{\alpha \sigma} \Theta^\rho_{\lambda \nu} \hat{P}^{\mu \sigma}_{\rho \beta} - \Theta^\alpha_{\alpha \sigma} \Theta^\rho_{\lambda \beta} \hat{P}^{\mu \sigma}_{\nu \rho} \nonumber \\  + \Theta^\sigma_{\alpha \lambda} \Theta^\rho_{\sigma \nu} \hat{P}^{\mu \alpha}_{\rho \beta} + \Theta^\sigma_{\alpha \lambda} \Theta^\rho_{\sigma \beta} \hat{P}^{\mu \alpha}_{\nu \rho} - \Theta^\sigma_{\alpha \lambda} \Theta^\mu_{\sigma \rho} \hat{P}^{\rho \alpha}_{\nu \beta} - \Theta^\sigma_{\alpha \lambda} \Theta^\alpha_{\sigma \rho} \hat{P}^{\mu \rho}_{\nu \beta} \nonumber \\  + \Theta^\sigma_{\alpha \nu} \Theta^\rho_{\lambda \sigma} \hat{P}^{\mu \alpha}_{\rho \beta} + \Theta^\sigma_{\alpha \nu} \Theta^\rho_{\lambda \beta} \hat{P}^{\mu \alpha}_{\sigma \rho} - \Theta^\sigma_{\alpha \nu} \Theta^\mu_{\lambda \rho} \hat{P}^{\rho \alpha}_{\sigma \beta} - \Theta^\sigma_{\alpha \nu} \Theta^\alpha_{\lambda \rho} \hat{P}^{\mu \rho}_{\sigma \beta} \nonumber \\  + \Theta^\sigma_{\alpha \beta} \Theta^\rho_{\lambda \nu} \hat{P}^{\mu \alpha}_{\rho \sigma} + \Theta^\sigma_{\alpha \beta} \Theta^\rho_{\lambda \sigma} \hat{P}^{\mu \alpha}_{\nu \rho} - \Theta^\sigma_{\alpha \beta} \Theta^\mu_{\lambda \rho} \hat{P}^{\rho \alpha}_{\nu \sigma} - \Theta^\sigma_{\alpha \beta} \Theta^\alpha_{\lambda \rho} \hat{P}^{\mu \rho}_{\nu \sigma} \,, \label{eqn:wu-double-covD-final}
\end{gather}
which is the final result. One then needs to replace $\Theta^\sigma_{\alpha \beta}$ and $\hat{P}^{\mu \rho}_{\nu \sigma}$ as given in the main text in Eqs.~\eqref{eqn:wu-connection-correction} and~\eqref{eqn:hat-P-hcg} respectively, in terms of the metric and scalar field of the scalar--tensor theory.

\bibliography{References}
\bibliographystyle{JHEP}

\end{document}